\documentstyle[preprint,aps]{revtex}
\font\bfslant=cmmib10

\begin{document}

\draft
\title{
Statistical Theory of Parity Nonconservation in Compound Nuclei}

\author{S.~Tomsovic}
\address{Dept. of Physics,
Washington State University,
Pullman, WA 99164}
\author{Mikkel B.~Johnson, A. C. Hayes, and J. D. Bowman}
\address{Los Alamos National Laboratory,
Los Alamos, NM  87545}

\date{\today}
\maketitle

\begin{abstract}
We present the first application of statistical spectroscopy to study the
root-mean-square value of the parity nonconserving (PNC)
interaction matrix element $M$ determined experimentally by scattering
longitudinally polarized neutrons from compound nuclei.  Our effective PNC
interaction consists of a standard two-body meson-exchange piece and a
``doorway" term to
account for $0^-$ spin-flip excitations.  Strength functions are calculated
using realistic single-particle energies and a residual strong interaction
adjusted to fit the experimental density of states for the targets, $^{238}$U
for $A\sim 230$ and $^{104,105,106,108}$Pd for $A\sim 100$.  Using the standard
Desplanques, Donoghue, and Holstein estimates of the weak PNC meson-nucleon
coupling constants, we find that $M$ is about a factor of 3 smaller than the
experimental value for $^{238}$U and about a factor of 1.7 smaller for Pd.
The significance of this result for refining the empirical determination of
the weak coupling constants is discussed.
\end{abstract}

\section{INTRODUCTION}

Studies of the weak interaction in strongly interacting systems have been of
great interest in nuclear physics for a number of reasons.  First of 
all, on the
experimental side, observables may be obtained cleanly from measurements
because: (1) the symmetries respected by the weak interaction are different
from those respected by the strong interaction, allowing the weak interaction
sector to be uniquely disentangled from the purely strong interaction sector;
and (2) the strength of the weak interaction is sufficiently small that its
matrix element between definite eigenstates of the strong interaction
Hamiltonian may be extracted perturbatively.  On the theoretical side,
knowledge of the weak interaction observables is a rich source of insights into
the behavior of many-body systems.  In those cases where the weak interaction
is known, the experimental observable provides an opportunity to learn about
the nuclear many-body wave function, giving an often unique glimpse at a
specific feature of the eigenstate of the nuclear Hamiltonian.  If, on the
other hand, the weak interaction is not completely known (as is the case for
the hadronic
weak interaction), the measurements offer the possibility of learning about
the weak interaction itself.  To realize this goal, the main theoretical task
is to calculate the properties of the nuclear many-body system as accurately as
possible with a Hamiltonian that is known to describe the initial and final
states in the matrix element.

Experimental information likely to shed light on the nature of the hadronic
weak interaction is available in both
light and heavy nuclei.  Measurements of weak-interaction observables in these
cases are driving the development of theoretical formulations of the many-body
problem along two principal lines distinguished by the significance placed on
the interpretability of individual matrix elements.  The shell model 
formulation
of the many-body problem is an example of a nonstatistical approach, and here
one envisions individual matrix elements, such as those of operators
describing transitions between discrete nuclear levels, to be calculable and
hence meaningful.  In light nuclei~\cite{adelberger} the data
have been available for many years, and these data are being analyzed using
such a shell-model approach.

Our particular interest is in heavy nuclei.  In this case~\cite{garvey},
measurements based on longitudinally polarized neutron scattering and focusing
on the hadronic parity nonconserving (PNC) interaction have
become available only in the last 5 - 10 years, and development of the
microscopic statistical approach is being stimulated by the availability of
this data.  The main focus of the latest experiments~\cite{triple} has
been the strength function of the effective PNC interaction in nuclei in the
mass $A\sim 230$ and mass $A\sim 100$ regions of the periodic table.  The
root-mean square (RMS) matrix element $M$ of this interaction is the physically
meaningful quantity, as evidenced by the fact that the nuclear spacing between
the resonances involved is on the order of 10 - 100 eV; note that it is not
possible to obtain individual many-body matrix elements from this data.  The
values of $M$ have been obtained, and these provide a new means to study the
parity violating interaction in nuclei based on statistical
theory~\cite{garvey}.
 
The particular formulation on which we base our paper is that of statistical
spectroscopy~\cite{french}.  Statistical spectroscopy has proven very 
successful
for calculations of strength functions of positive parity operators, but it has
not been developed as fully for negative parity operators, especially for
the case of parity violation.  In this paper we will develop the theory in a
study of parity violation for the targets measured in Ref.~\cite{triple},
namely $^{238}$U for $A\sim 230$ and $^{104,105,106,108}$Pd for $A\sim 100$.
The statistical densities that we will need are calculated using
realistic single-particle
energies and a residual strong interaction adjusted to fit the experimental
density of states in the two elements of interest here.  The
determination of the smoothed strength and level densities is well established,
and the theory has been shown to be well justified through numerical and
theoretical evaluations in the early work~\cite{french}.  The status of
statistical spectroscopy in the other nuclei measured in Ref.~\cite{triple}
will be explored more fully in a later paper~\cite{st}.  On the other hand,
the application of statistical spectroscopy to parity violation raises new
issues, and it is the main purpose of this paper to identify the new
issues and show how statistical spectroscopy may be implemented in view of
them.

We discuss the statistical observables in Sect.~II, emphasizing their
advantage for study of the weak PNC interaction in compound nuclei.
In Sect. III and the Appendix, we emphasize the microscopic foundation of
the statistical theory and also collect ingredients that will be used
in this paper for specific calculations.  The limited aim of our
calculations of $M$, presented in Sect.~IV, is to benchmark the theory, i.e.
provide reference points for future calculations based on standard $\pi$ plus
$\rho$ meson exchange interactions with strengths taken from the existing
literature (specifically we use the ``DDH best" set, see
Ref.~\cite{adelberger}).
Since we are not adjusting parameters to fit data, we leave open the
question of whether the data prefers some alternative parameter
set~\cite{st}.

In Sect.~V we discuss sources of uncertainty in such a
calculation and indicate additional studies that would establish a more
refined theoretical error analysis.  Finally, in Sect.~VI we present a summary
of our paper and draw our conclusions.

\section{STATISTICAL AND NONSTATISTICAL LIMITS}

Experiments are often able to determine individual matrix elements
$<\Psi_I|O|\Psi_K>$ of some operator $O$ between stationary states $|\Psi_I>$
and $|\Psi_K>$ of the nucleus.  Microscopic, nonstatistical calculations
of such matrix elements are often quite successful.  However, such calculations
of individual matrix elements are not guaranteed to be stable.  For example,
a basic problem arises in calculating weak interaction operators $O$ (e.g.,
the axial charge operator or the hadronic PNC interaction)
within the shell-model framework.  This is because the matrix elements,
\begin{equation}
<\Psi_I|O|\Psi_K>=\sum_{i,k}a^{*i}_I<\phi_i|O|\phi_k>a^{k}_K ,
\label{eq:2}
\end{equation}
which we have expressed in terms of the expansion coefficients $a^i_I$ of the
nuclear eigenstates $|\Psi_I>$ in a many-body basis $|\phi_i>$,
\begin{equation}
|\Psi_I>=\sum_{i}a^{i}_I|\phi_i> ,
\label{eq:1}
\end{equation}
often involve strong cancellations so the predictions are sensitive to small
changes in the Hamiltonian of the strong interaction.

It is instructive to see how this sensitive interplay comes about.  In order
to avoid the problem of diagonalizing the complete many-body Hamiltonian,
one assumes that the important pieces of the nuclear wave functions are the
low-lying components, which one isolates by making a truncation into a ``model
space".  Because the components in the model space correspond to just a
piece of the wave function, one must make a correction, either by applying a
renormalization to the hadronic weak interaction to account for the excitations
out of the model space not accounted for by the effective interaction, or by
enlarging the model space.  In the shell-model theory of Ref.~\cite{haxton1},
this was done by enlarging the space.  Enlarging the space was accompanied
by corresponding changes in the coefficients $a^{i}_I$ and $a^{k}_K$,
leading to dramatic changes in the matrix elements, illustrated very clearly in
Ref.~\cite{haxton1}.  It is the conclusion of this and other investigations
that because of the sensitive interplay among various components, the
nuclear wave function must be calculated very precisely, especially by
including multiple particle-hole (p-h) admixtures and carefully eliminating
spurious center-of-mass motion.  To avoid this strong model dependence,
it was pointed out that the PNC matrix element could be calculated from the
measured analog first forbidden decay of $^{18}$Ne.

With more computer power the individual matrix elements may of course be
calculated in light nuclei with sufficient stability.  However for
heavy nuclei, where one has many more degrees of freedom, chaotic behavior
may occur, and then the problem becomes much more serious.  In this case
eigenstates of the
Hamiltonian have so many components, with amplitudes roughly equal in size and
with phases that occur in a nearly random fashion, that the individual matrix
elements become impossible to interpret in principle.  Even small changes in
the Hamiltonian (or approximation scheme) can dramatically affect the wave
function and hence the expectation value of an observable~\cite{bohigas}.
(The effective dimensionality, or number of significant Slater determinant
components in the eigenstates, is far more stable.)  Although a large amount
of information is irretrievably lost in the transition from simple to complex
systems in this fashion, significant physically meaningful information does
remain.  Indeed, it was the original motivation in developing the theory of
statistical spectroscopy to identify this information and develop methods
to extract it.  The chaotic behavior of the nuclear wave function suggests
application of the central limit theorem for this purpose and leads to the
notion that matrix elements of operators may be thought of as elements of a
statistical distribution.

For chaotic nuclei, one is thus led to identify as the observable
the square of the matrix element of the PNC interaction averaged over $N_I$ and
$N_K$ states of appropriate symmetry occurring in a small window of energy,
\begin{equation}
M^2\equiv{1\over N_IN_K}\sum_{I,K}|<\Psi_I|O|\Psi_K>|^2,
\label{eq:3}
\end{equation}
which can be cast into a form very similar to Eq.~(\ref{eq:2}) using the
statistical
properties of the wave function.  To see this, first use Eq.~(\ref{eq:1}) to
rewrite
Eq.~(\ref{eq:3}) as follows,
\begin{equation}
M^2=\sum_{i_1,k_1}\sum_{i_2,k_2}({1\over N_I}\sum_{I}a^{*i_2}_Ia^{i_1}_I)
<\phi_{i_1}|O|\phi_{k_1}><\phi_{i_2}|O|\phi_{k_2}>^*
({1\over N_K}\sum_{K}a^{*k_2}_Ka^{k_1}_K).
\label{eq:4}
\end{equation}
Since the phases of the components are random when the statistical approach
is valid, we may write
\begin{equation}
{1\over N_K}\sum_{K}a^{k_2*}_Ka^{k_1}_K=\delta(k_1,k_2)
{1\over N_K}\sum_{K}|a^{k_2}_K|^2 ,
\label{eq:5}
\end{equation}
and if the $a^k_K$ ($1\leq k\leq n$, where $n$ is the effective dimensionality)
in the averaging interval are additionally
assumed to be elements of an ensemble of Gaussian distributed random variables,
we may write for each $k_2$
\begin{equation}
{1\over N_K}\sum_{K}|a^{k_2}_K|^2={1\over n} .
\label{eq:6}
\end{equation}
Typically, $N_I$ and $N_K$ are rather small in comparison to the effective
dimensionality $n$.  Equation
(\ref{eq:6}) is the stronger assumption.  In Ref.~\cite{tomsovic}, the space is
partitioned with the different partitions being described by separate
ensembles.  If we adopt the weaker statistical assumption of Eq.~(\ref{eq:5}),
Eq.~(\ref{eq:4}) becomes
\begin{equation}
M^2={1\over N_IN_K}\sum_{I,K}\sum_{i,k}|a^{*i}_I<\phi_i|O|\phi_k>a^k_K|^2
\label{eq:7}
\end{equation}
Comparing Eq.~(\ref{eq:7}) to Eq.~(\ref{eq:2}), we see that now the different
terms in the sum
over $i,k$ occur with the ${\it same}$ sign, so that the problem with delicate
cancellations that would prevent a meaningful calculation of an individual
matrix element is avoided.  Thus,
observables amenable to a statistical analysis are robust against
approximations that do not change the overall scale characterizing the
distribution of the $a^i_I$.

In actual situations where the statistical theory is applicable, only the
low-lying components $a_{K}^k$ of the many-body wave function are statistical
in the sense of Eq.~(\ref{eq:5}).  We look at a more systematic method for
separating the statistical and nonstatistical behavior in Sect.~III  below,
and find that when a more careful separation is made, we arrive at the same
conclusion.

\section{FORMULATION OF STATISTICAL THEORY FOR PARITY VIOLATION}

The specific problem we will be addressing in trying to understand the
TRIPLE data of Ref.~\cite{triple} is how to calculate $M$ when the operator
$O$ in Eq.~(\ref{eq:7}) is the interaction $V^{PNC}$, corresponding to the
hadronic parity-violating interaction between two particles in free-space;
we base our numerical results in this paper on a meson-exchange form
for the weak PNC interaction with the coupling strengths presented
by Desplanques, Donoghue, and Holstein~\cite{ddh} (DDH).  Although these
coupling strengths were estimated by DDH from empirical non-leptonic
hyperon decay data, they also incorporated some estimates of various 
contributions such as the strong interaction enhancements to the symmetry 
breaking; as a result, the couplings still contain a considerable amount of 
uncertainty.  Because of these 
uncertainites and the fact that the DDH interaction has not been fitted to 
nuclear PNC data, one does not expect to be able to explain the TRIPLE 
measurements without some further refinement of the coupling parameters.

In our formulation, we will be using the fact that statistical
methods~\cite{french} provide a relatively straightforward and
model-independent means to average over the important components $(i,k)$ in
Eq.~(\ref{eq:7}), namely those that mix chaotically among themselves by the
action of the strong interaction.  These components correspond to the
independent-particle configurations formed by placing valence particles in
single-particle orbitals that lie within a few MeV of each other (within the
spreading width) outside a closed shell, in other words, to components within
a model space.  The central limit theorem is used to express the
strength functions in terms of bivariate Gaussian functions characterized by a
few parameters as described in Sect.~III.A below.  Statistical mixing of
single-particle states lying farther than a typical spreading width (a few
MeV) away from these is weak~\cite{fk83} and naturally suppressed in the
statistical formulation.

However, statistical mixing is unfortunately only part of the problem of
understanding the TRIPLE data.  Nonstatistical mixing of states lying outside
the model space is expected to occur (see Sect.~III.B.3), and the real
question at issue in applying statistical spectroscopy to parity violation is
how to deal in a satisfactory way with important components of the wave
function that lie just $1\hbar\omega$ outside the model space, for example
the giant resonances.  One solution to this problem is provided by effective
interaction theory~\cite{ellis}.

Effective interaction theory provides book-keeping techniques for building the
contribution of the $(i,k)$ components lying outside the model space in
Eq.~(\ref{eq:7}) into an ${\it effective}$ operator $\overline{O}$ acting
entirely within the model space.  More generally, it prescribes not only how
one can calculate microscopically various effective operators $\overline{O}$
corresponding to actual observables $O$, but also the effective Hamiltonian
$\overline{H}$ whose eigenfunctions $|\overline{\Psi}_K>$,
\begin{equation}
\overline{H}|\overline{\Psi}_K>=E_K|\overline{\Psi}_K> ,
\label{eq:8}
\end{equation}
correspond to the true stationary states $|\Psi_K>$.  Once the model space has
been set up, matrix elements of the Hermitian, energy-independent operators
$\overline{H}$ and $\overline{O}$ may be calculated diagrammatically according
to the rules of degenerate many-body perturbation theory~\cite{john2}.  The
connection to the observable given in Eq.~(\ref{eq:2}) is
\begin{equation}
<\overline{\Psi}_I|\overline{O}|\overline{\Psi}_K>=<\Psi_I|O|\Psi_K> .
\label{eq:10}
\end{equation}
An application of effective interaction theory for the calculation of
PV observables in the sense outlined here was followed in Ref.~\cite{john1}.

Once one has obtained the effective operator $\overline{V}^{PNC}$
corresponding to $V^{PNC}$ by making use of this theory, one may apply the
powerful methods of statistical spectroscopy directly to $\overline{V}^{PNC}$.
The expression for $M^2$ in Eq.~(\ref{eq:7}) may be expressed in the model
space by expanding $|\overline{\Psi}_K>$ as in Eq.~(\ref{eq:1}) to obtain
\begin{equation}
|\overline{\Psi}_K>=\sum_{k}\overline{a}^k_K|\phi_k> ,
\label{eq:13}
\end{equation}
where the sum now runs over independent-particle configurations $k$ lying
purely in the model space.  (Note that the coefficients $\overline{a}^{k}_K$ do
not necessarily have a simple connection to $a^{k}_K$.)  By substituting
Eqs.~(\ref{eq:10}) and (\ref{eq:13}) into Eq.~(\ref{eq:2}), we find
\begin{equation}
<\Psi_I|O|\Psi_K>=<\overline{\Psi}_I|\overline{O}|\overline{\Psi}_K>
=\sum_{i,k}\overline{a}^{*i}_I<\phi_i|\overline{O}|\phi_k>\overline{a}^{k}_K .
\label{eq:14}
\end{equation}
Similarly, the model space counterpart to Eq.~(\ref{eq:7}) is found by
substituting Eqs.~(\ref{eq:14}) into Eq.~(\ref{eq:3}) and following the
analogous reasoning leading from Eqs.~(\ref{eq:4}) to Eq.~(\ref{eq:7}),
\begin{equation}
M^2={1\over N_IN_K}\sum_{I,K}\sum_{i,k}|\overline{a}^{*i}_I
<\phi_i|\overline{O}|\phi_k>\overline{a}^k_K|^2 .
\label{eq:15}
\end{equation}
The connection between $M$ in Eq.~(\ref{eq:15}) measured in a region of
excitation energy $E$, the effective parity-violating interaction, and the
strength-function $S_{\overline{O}}$ defined below
in Sect.~III.A (with $\overline{O}
= \overline{V}^{PNC}$) is
\begin{eqnarray}
\label{stat}
M^2\equiv M^2(E_1,E_2,\Gamma_1,\Gamma_2)=D(E_1,\Gamma_1)D(E_2,\Gamma_2)
S_{\overline{O}}(E_1,\Gamma_1,E_2,\Gamma_2) ,
\end{eqnarray}
where $D(E,\Gamma)$ is the mean spacing between neutron resonances and
$\Gamma$ is the set of conserved quantum numbers labeling the states of
interest (i.e., angular momentum, parity, isospin, etc.).

As we have emphasized, in Eqs.~(\ref{eq:14}) and (\ref{eq:15}) the sums run
only over independent-particle configurations $i,k$ lying within the model
space.  By comparing these equations we arrive at a conclusion similar to that
found at the end of Sect.~II, namely that the delicate cancellations that
occur among amplitudes
$\overline{a}^{*i}_I <\phi_i|\overline{O}|\phi_k>\overline{a}^k_K$ in
calculations of individual matrix elements are absent when one calculates the
mean-square matrix elements in effective interaction theory.  However, we also
see that some residual interferences may creep back into the statistical
problem through the calculation of $\overline{O}$, whose various terms may
have fluctuating signs.  Since these signs do not arise from statistical
behavior, one may expect that this method of calculating mean-square matrix
elements is the more robust procedure when statistical behavior is indicated.

Note how the statistical and nonstatistical aspects of the many-body physics
cooperate to produce physical observables in this formulation:  the
statistical behavior is confined to the model space and handled within
statistical spectroscopy, whereas the nonstatistical behavior (including any
$1\hbar\omega$ collectivity that may arise from giant resonances, for example)
is concentrated in the effective operator terms $\overline{O}-O$.  These two
aspects are interwoven at the end when $M$ is evaluated in Eq.~(\ref{eq:15})
or Eq.~(\ref{stat}).

\subsection{Statistical Spectroscopy of French and Collaborators}

 From a statistical point of view, spectroscopy in many-body systems reduces to
two largely independent physical features, fluctuations, and second, 
the secular
behavior of system dependent properties such as level densities, expectation
values, and strength densities.  Random matrix theories, first introduced by
Wigner~\cite{wigner}, and its growing number of generalizations especially
with regards to mesoscopic, many-body electron systems~\cite{altshuler},
address the physics of fluctuation properties.  This is the most widely
recognized aspect of a statistical approach, but is not our concern here.
Instead, we shall only describe an implementation of statistical spectroscopy
that describes system-dependent large-scale secular behaviors that arise
in the construction of the strength function of
$\overline{V}^{PNC}$ in
chaotic regimes of heavy nuclei.  Just as Weyl~\cite{weyl} was able to show
that to leading order in wave number the only surviving information 
in the count
of modes in a cavity was its volume, generalized central limit theorems (CLTs)
and moment methods, as developed by French and co-workers over the past thirty
years~\cite{french}, identify the analogous surviving information in a
microscopic approach to many-body spectroscopy and provide the means to
calculate secular behaviors.  In this context, convergence of the
CLTs improves with the dilute limit of many valence particles occupying a much
larger number of available single particle states.  It should be 
noted also that
experience has shown good convergence to the CLT forms even in ds shell model
examples~\cite{french} that involve spaces much smaller and less dilute than
those appropriate for medium and heavy nuclei.

The matrix element $M$ defined in Eq.~(\ref{eq:3}) is an example of a  strength
density, $S_{\overline{O}}(E_1,\Gamma_1;E_2,\Gamma_2)$, which may be expressed
as a
trace.  In statistical spectroscopy, we consider the trace decomposed
into its sum over particle configurations, ${\bf M}=(m_1,m_2,m_3,...)$ where
each $m_i$ is the number of particles in a subset of single particle levels and
${\bf N}=(N_1,N_2,N_3,...)$.  The grouping is not restricted to a particular
scheme, but for the present discussion each subset is taken to be a j-orbital.

In general, strength densities are bivariate functions of energy.  Using mainly
the notation of French et al.~\cite{french},
i.e.~$\langle\langle~\rangle\rangle$ indicates the trace and $\langle~\rangle$
indicates the expectation value ($\langle\langle~\rangle\rangle / N$),
the expression is
\begin{equation}
\label{strengthdensity}
S_{\overline{O}}(E_1,\Gamma_1;E_2,\Gamma_2)=\langle\langle {\overline{O}}
^\dagger P_{\Gamma_2} \delta(E_2
- \overline{H})  {\overline{O}} P_{\Gamma_1} \delta(E_1-\overline{H}
) \rangle\rangle
\end{equation}
where $\overline{H}$ is the effective nuclear Hamiltonian, $P_{\Gamma}$ is
the projector onto the $\Gamma$ subspace, and $\overline{O}$ is, for example,
the effective parity violating interaction.  It is useful to  perform a
partitioning decomposition of Eq.~(\ref{strengthdensity});
\begin{eqnarray}
\label{sdpart}
S_{\overline{O}}(E_1,\Gamma_1;E_2,\Gamma_2)&=&\sum_{{\bf M},{\bf
M^{\prime}}}\
\langle\langle \overline{O}^\dagger P_{\Gamma_2}\delta(E_2 - \overline{H})
P_{\bf M^{\prime}} \overline{O}
P_{\Gamma_1}
\delta(E_1-\overline{H})P_{\bf M}\rangle\rangle \nonumber \\
&=&\sum_{{\bf M},{\bf M^{\prime}}}\ S_{\overline{O}}
(E_1,\Gamma_1;E_2,\Gamma_2;{\bf M},{\bf M^\prime})
\end{eqnarray}
Parity is taken care of trivially since the configurations respect it
and one can just restrict the ${\bf M}$, ${\bf M}^\prime$ sums to the
appropriate parities.  For the model spaces of the medium and heavy
nuclei considered in this paper, only the angular momentum
decomposition remains to be taken into account.
If $\overline{O}$ were an electromagnetic transition
operator that coupled different angular momenta, we would have to modify the
angular momentum decomposition given below.  However for the case we are
interested in here, $\overline{O}$ is a parity violating interaction and is a
$J$-scalar.  In Eq.~(\ref{strengthdensity}), $P_{\Gamma_2}$ commutes with
$\overline{O},\overline{H}$ and can be translated next to $P_{\Gamma_1}$.  The
strength expression vanishes unless $\Gamma_2= \Gamma_1=\Gamma$.  There are
two additional simplifications.  For the purposes of the angular momentum
decomposition relevant to the parity violation calculations, $E_1=E_2=E$.
Finally, as a consequence of statistical spectroscopy, the spreading widths
are nearly constant.  Therefore, $S_{\overline{O}}(E,E,\Gamma)$ is proportional
to the J-decomposed density of states, where the constant of proportionality
$C(j|E)$ is the conditional probability density of finding the angular momentum
$j$ given $E$.  We find,
\begin{eqnarray}
\label{sdbethe}
S_{\overline{O}}(E_1,\Gamma_1;E_2,\Gamma_2)&=&S_{\overline{O}}
(E_1,E_2;\Gamma) \nonumber \\
&\approx&C(j|E)  S_{\overline{O}}(E,E)\nonumber \\
&=&C(j|E) \sum_{{\bf M},{\bf M^{\prime}}} S_{\overline{O}}
(E,E;{\bf M},{\bf M^\prime})
\end{eqnarray}
To the level of our statistical approximations, $C(j|E)$ is well approximated
by a partitioned version of Bethe's form~\cite{bethe} with an energy-dependent
spin cut-off factor, $\sigma^2_j(E,{\bf M})$, evaluated as
\begin{eqnarray}
\label{bethe}
C(j|E)&=& \left({\sum{N_i} \atop \sum{m_i}}\right)^{-1}\sum_{\bf M}
C(J|E;{\bf M})\prod_{i}\left({N_i \atop m_i}\right)\nonumber
\\C\left(j|E;{\bf M}\right)&=&\left({(2j+1)^2 \over 8\pi
\sigma^6_j(E,{\bf M})}
\right)^{1/2}\exp\left(-{(2j+1)^2\over 8\sigma^2_j(E,{\bf M})}\right) \nonumber
\\ 3\sigma^2_j(E,{\bf M})&\approx& \langle J^2\rangle^{\bf M}+ \langle J^2 (H
-E_{\bf M})\rangle^{\bf M}\ {E-E_{\bf M}\over \sigma^2_{\bf M}}+\ ...
\end{eqnarray}
where
\begin{equation}
\label{moment}
E_{\bf M} = \sum_i m_i \epsilon_i
\end{equation}
and
\begin{equation}
\label{moments}
  \sigma^2_{\bf M}= \langle V^2\rangle^{\bf M}=\sum_{r\leq s
\atop t \leq u} {(N_r-m_r)(N_s-m_s-\delta_{rs})m_t(m_u-\delta_{tu}) \over
(N_r-\delta_{rt}-\delta_{ru})(N_s-\delta_{st}-\delta_{su}-\delta_{rs}) N_t
(N_u-\delta_{tu})}\sum_\Gamma [\Gamma]|V^\Gamma_{rstu}|^2 \nonumber \\
\end{equation}
(below we define $V^\Gamma_{rstu}$)
with $\epsilon_i$ the single particle energy of the $i^{th}$ orbital including
any contribution to the energy from the part of the two-body interaction which
transforms like a one-body operator.  It is further supposed that the single
particle basis diagonalizes the full one-body part of the Hamiltonian.  The
$V^\Gamma_{rstu}$ are the matrix elements of the remaining two-body
interaction.  The counting factor, $[\Gamma]$, reduces to $2j+1$ above.

In addition, only those terms survive in Eq.~(\ref{sdbethe}) in which
the fundamental matrix elements of $\overline{O}
$ appear as absolute squares.  We thus have
\begin{equation}
\label{sdpart2}
S_{\overline{O}}(E_1,E_2;{\bf M},{\bf M^\prime})= \prod_{i,j} \left({N_i \atop
m_i}\right)
\left({N_j \atop m_j}\right) |\langle {\bf M^\prime}|\overline{O}
| {\bf M}\rangle|^2
\rho(E_1,E_2;{\bf M},{\bf M^\prime})
\end{equation}
to evaluate.  If $\overline{O}$ is
a two-body operator, the matrix element can be
deduced from Eq.~(\ref{moments}) with the $\{r,s,t,u\}$ sums suppressed and
if not, from the appropriate analogous equations for other rank operators.
The theoretical expression for $\rho(E_1,E_2;{\bf M},{\bf M^\prime})$, which
can be derived assuming statistical properties of many-body wave functions and
has been confirmed in numerical tests, is that of a unit normalized
bivariate Gaussian,
\begin{eqnarray}
\label{bivar}
\rho(E_1,E_2;{\bf M},{\bf M^\prime})&=&{1\over 2\pi \sigma_{\bf M} \sigma_{\bf
M^\prime}\sqrt{1-\xi^2}} \exp\left(-{1\over 2(1-\xi^2)}\left\{ {(E_1-E_{\bf
M})^2 \over \sigma_{\bf M}^2}\right. \right. \nonumber\\
&~& \left. \left. - { 2\xi (E_1-E_{\bf
M})(E_2-E_{\bf M^\prime})
\over \sigma_{\bf M} \sigma_{\bf M^\prime}} + {(E_2-E_{\bf M^\prime})^2 \over
\sigma_{\bf M^\prime}^2 } \right\}\right)
\end{eqnarray}
which to specify completely requires the evaluation of two centroids, $E_{\bf
M},E_{\bf M^\prime}$, two variances $\sigma_{\bf M}^2, \sigma_{\bf M^\prime}^2$
and a normalized correlation coefficient, $\xi$.  In principle, these moments
are given by
\begin{eqnarray}
\label{sdmoments}
S_{00}&=&\langle\langle \overline{O}^\dagger P_{\bf M^\prime} \overline{O}
  P_{\bf M} \rangle\rangle
~~~~~~~~~~~~~~ \qquad \xi=\langle\langle \overline{O}
^\dagger \overline{H}P_{\bf M^\prime} \overline{O}\overline{H}P_{\bf
M}
\rangle\rangle/(S_{00} \sigma_{\bf M} \sigma_{\bf M^\prime}) \nonumber \\
E_{\bf M}&=&\langle\langle \overline{O}^\dagger P_{\bf M^\prime}\overline{O}
\overline{H} P_{\bf M}
\rangle\rangle/S_{00} ~
\qquad E_{\bf M^\prime}=\langle\langle \overline{O}^\dagger \overline{H}
  P_{\bf M^\prime} \overline{O} P_{\bf M}
\rangle\rangle/S_{00} \nonumber \\
\sigma^2_{\bf M}&=&\langle\langle \overline{O}
^\dagger P_{\bf M^\prime}\overline{O}\overline{H}^2 P_{\bf
M}\rangle\rangle/S_{00}
\qquad \sigma^2_{\bf M^\prime}=\langle\langle \overline{O}^\dagger \overline{H}
^2 P_{\bf M^\prime} \overline{O}
P_{\bf M} \rangle\rangle/S_{00} \nonumber \\
\end{eqnarray}
However, the centroids and variances can be replaced by the simpler expressions
given in Eqs.~(\ref{moment},\ref{moments}) with almost no loss in accuracy of
the approximation.

The correlation coefficient $\xi$ is quite important.  As it approaches zero,
strength is distributed uniformly in the sense that the many-body mean square
matrix element is a constant independent of $(E_1,E_2)$.  In the opposite limit
of $\xi\rightarrow 1$, the mean square matrix elements of $\overline{O}$ are
concentrated
along the diagonal and their behavior approaches being inversely 
proportional to
the level density; see ahead.  Although, $\xi$ depends on $(\bf M,
\bf M^\prime)$, the fluctuations from configuration to configuration
are small and the errors made in using a single averaged value negligible
compared to other approximations.  In fact, we use Gaussian ensembles with
operators of fixed  particle rank to evaluate $\xi$ since the strength density
is not extremely  sensitive to small variations.  In the dilute limit of
$N$ (number of single-particle states) $\rightarrow \infty$,
$\overline{H}$ a $k$-body operator, $\overline{O}$ a $k^\prime$-body
operator, and $m$ valence
particles,
\begin{equation}
\xi=\left({m-k^\prime \over k}\right)/\left({m \over k}\right)\approx
1-kk^\prime/m
\end{equation}
The finite $N$ expression is more complicated but analytically known for a
Gaussian ensemble.  It is this expression that we use and which can be found in
Ref.~\cite{tomsovic}.  It is not sufficiently illuminating to
give the expression here. Typical values for $\xi$ in medium and heavy nuclei
are $\sim 0.7 - 0.9$.

The criterion for choosing an appropriate model space is more 
complicated in the
strength forms than for the level density since it depends on the joint
contribution of $({\bf M}, {\bf M^\prime})$, and the matrix elements connecting
configurations. Nevertheless, one uses a similar logic and the most important
configurations are the same as in the level density calculation, i.e. the
single-particle orbitals strongly mixed by the residual two-body interaction.
If ${O}$ connects to important configurations outside the model space,
statistical methods must be extended to incorporate perturbative
effects as discussed in the following subsection.

\subsection{Model Space and Effective Interactions}

Having made these general remarks, we next discuss the choice of model space
and present the contributions to the effective one-body and two-body
parity-violating interactions that we will use in this work.
In the theory, there exist also three- and higher-body contributions
to the effective operator $\overline{O}$, but there have been no calculations
of these terms so we will omit them.

\subsubsection{Model Space and Strong Interaction}

There are two practical issues with regard to the choice of the strong
interaction in statistical spectroscopy.  One is the specification of the
one-body part of the interaction, which specifies the model space and single
particle energies.  The other is the choice of the spreading interaction that
is crucial for getting the level density correct.  Once these two elements
are specified, then the partitioning of the model space and the calculation
of the bivariate densities in Sect.~III.A are completely specified.

By ``spreading interaction", one means that part of $\overline{H}$ that
spreads the configurations across the eigenstates.  To determine it from a
given $\overline{H}$ requires
removing those parts responsible for shifting the centroids of levels.  In
the language of Ref.~\cite{french}, the spreading interaction is the
irreducible rank-2 part of $\overline{H}$.  The shell-model or mean-field
contribution, which determines the single-particle spectrum,
is the rank-one contribution of $\overline{H}$.
Note that a somewhat different definition of one- and two-body
parts of effective interaction is used in microscopic shell-model studies,
where the single-particle energies are often taken to be those appropriate to
the beginning or end of the shell.  The residual interaction for multiple
numbers of particles in a shell can then both spread levels and shift their
centroids.  One must keep this distinction in mind when comparing the
shell-model and statistical approaches to avoid confusion.

In our applications of statistical spectroscopy, we choose both the mean-field
and spreading interactions phenomenologically.  By so doing, we thus depart
from a purely theoretical implementation of the theory.  The mean-field part is
identified with the shell-model potential.  By taking the single-particle
energies to be those corresponding to such a phenomenological potential, a
smooth $A$ dependence is imposed.  The spreading interaction is
adjusted phenomenologically to reproduce
the level density in the excitation region of nuclei of interest where the
weak spreading width is to be calculated.  The energy dependence of the level
densities for nuclei measured by TRIPLE have been extensively studied and
experimentally tabulated~\cite{ld},
and improved level densities have been obtained also in the region of interest
for $M$ in Ref.~\cite{triple}.  We used these results for our 
numerical studies.

The choice of model space is determined mainly by the considerations discussed
in Sect.~III, namely that the chaotic mixing occur within the model space.
Since the strong spreading width of states is on the order of 2 MeV, a model
space spanned by single particle levels lying within 2 or 3 MeV of each other
is expected to be sufficient for those nuclei in which there are at least
two or three particles (or holes) in the valence space for both neutrons and
protons.

For the nuclei of interest in this paper, it is a relatively
straightforward matter to choose the model space subject to these criteria
by looking at shell-model level schemes.  In spherical nuclei, the
single-particle orbitals will correspond
to eigenstates in a spherical potential, and in deformed nuclei to Nilsson
levels.  Since statistical spectroscopy has always been applied (even in
regions of deformation, for example the $A\sim 230$ region) exclusively in the
spherical basis, so we will do our present calculations in this basis as well,
in accord with our goal to benchmark the theory.  We estimate the importance
of deformation by including it through an effective one-body PNC interaction
(see Sect.~V.B).

In our work, the single-particle energies of the active orbits for $^{239}$U,
given in Table~I were taken from Ref.~\cite{tomsovic} and for Pd were
calculated~\cite{sp} using a Woods-Saxon potential,
\begin{eqnarray}
U=Vf(r)+V_{ls}{\bf \ell}\cdot{\bf s}{r_0^2\over r}{d\over dr}f(r)  \nonumber
\\f(r)=[1+exp({r-R\over a})]^{-1} ,
\label{eq:ws1}
\end{eqnarray}
with realistic parameters~\cite{bm}
\begin{eqnarray}
R=r_0 A^{1/3},~~~~~~r_0=1.27~{\rm fm},~~~~~~a=0.67~{\rm fm},  \nonumber
\\V=(-51 +33\tau_z{N-Z\over A})~{\rm MeV},~~~~~~V_{ls}=-0.44V .
\label{eq:ws2}
\end{eqnarray}
The resulting energies for the daughter $^{107}$Pd is given in Table~II.
The model space is
spanned by the set of all such configurations ($N$ in number) that can be
formed by putting $m$ valence particles in a few (active) single-particle
orbits lying in the immediate neighborhood of the nearest closed neutron and
proton subshells (we refer to this as a $0\hbar\omega$ space).  Note that
there are both positive and negative parity levels included.  The appearance
of the intruder states of opposite parity is a feature particular to heavy
nuclei:  for the light nuclei for which parity violation has been considered,
there are no such intruder states.

In the following two subsections, Sect.~III.B.2 and III.B.3, we discuss three
different pieces of $\overline{V}^{PNC}$, a standard one-body piece, a standard
two-body piece, and a higher-order correction.  The latter will be seen to be
peculiar to heavy nuclei, where the relative importance of the standard
one-body and two-body PNC interactions is different from what it is
in light nuclei.

\subsubsection{Lowest-Order Effective One- and Two-Body PNC Potentials}

The standard two-body piece of the effective PNC interaction in the
nucleus $V^{PNC(2)}_{Std}$ that we use is related to the free-space two-body
PNC interaction as discussed in Ref.~\cite{adelberger} and in the Appendix.
The standard one-body interaction $V^{PNC(1)}_{Std}$ is, in physical terms,
that part of the effective PNC interaction acting on nucleons outside
the core (the valence nucleons) that originates from the core nucleons.
Since $V^{PNC(1)}_{Std}$ is just an average of $V^{PNC(2)}_{Std}$ over the
core of the nucleus~\cite{adelberger}, we have a relatively simple connection
between the two.  Referring to the Fermi gas model (see Eq.~(20) of
Ref.~\cite{adelberger}), we obtain
\begin{equation}
V^{PNC(1)}_{Std}={\rho \over \rho_0}\sum_i(C_0+C_1 \tau_{zi})
{\bf \sigma_i\cdot p_i} ,
\label{eq:16}
\end{equation}
where $\rho$ is the density of the core and $\rho_0$ is the central density
in nuclei.  The coefficients $C_0$ and $C_1$ are given in Eq.~(\ref{eq:17}) of
the Appendix.  For
$V^{PNC(1)}_{Std}$ as well as $V^{PNC(2)}_{Std}$ we retain only the
contributions from $F_\pi$ and $F_0$, the weak $\pi$-NN and (isoscalar)
$\rho$-NN coupling constants, respectively, since these are the dominant
contributions from the DDH parameter set.

Next, we want to discuss the higher-order pieces of this interaction.
The situation can become quite involved, since
higher-order pieces of the effective one-body and two-body parity violating
interactions can be formed from both $V^{PNC(1)}_{Std}$ and
$V^{PNC(2)}_{Std}$.  It is our aim in this paper to limit attention to
issues that have been discussed already in the literature, so we
will not enumerate the other terms in any systematic fashion.  This limits us
to only one type of higher-order term, discussed in the next subsection.

\subsubsection{Higher-Order Contributions Arising from $V^{PNC(1)}_{Std}$}

One of the differences between parity violation in light and heavy nuclei is 
the relative importance of single-particle transitions in PNC observables.  
Large shell-model calculations (see for example Ref.~\cite{haxton2,brown})
indicate that the PNC transitions in light nuclei are largely single-particle 
in nature~\cite{adelberger,hayes}.  However in heavy nuclei, because of the 
appearance of intruder states of opposite parity in the $0\hbar\omega$ model 
space, $\overline{V}^{PNC(2)}$ can connect states within this space, and 
many-body mixing plays a much more important role.  The fact that there are 
generally no active single-particle states that are coupled by 
$V^{PNC(1)}_{Std}$ (i.e., that there are no opposite-parity single particle 
levels with the same total angular momentum inside the $0\hbar\omega$ model 
space in a heavy spherical nucleus, even though there are intruder states) 
means that $V_{Std}^{PNC(1)}$ acts indirectly.  Its contribution to the 
effective PNC interaction may be enumerated in second- and 
higher-order perturbation theory; in the theory of this paper, these are the
major contributors to the difference of the operators $O-\overline{O}$ defined
in Eq.~(\ref{eq:10}) and accounts fully for the effect of $V^{PNC(1)}_{Std}$
on $M$.  In a deformed nucleus the situation is somewhat different, 
see Sect. V.

The leading perturbative contribution of $V^{PNC(1)}_{Std}$ to
$\overline{V}^{PNC}$ has been discussed in Ref.~\cite{flam}.  In an effort
to use physical ideas to identify the important correction,
Auerbach~\cite{auer} developed the idea of ``doorway" states for the PNC
interaction.  In this approach, the mixing of $V^{PNC(1)}_{Std}$ is mediated
by the isoscalar and isovector $0^-$ spin-flip giant resonances.  The
disadvantage of Auerbach's formulation is that it is highly phenomenological.
A microscopic expression $V^{PNC(2)}_{Dwy}$ for the doorway contribution to
the effective two-body PNC interaction was obtained in Ref.~\cite{john1} by
identifying the doorway state as a collective one-particle one-hole $0^-$
phonon in the Tamm-Dancoff approximation (TDA).  The result
brought together the results of Refs.~\cite{flam,auer}, showing that the
effect of the doorway is to renormalize the perturbative contribution of
Ref.~\cite{flam} by a factor $(\omega_0/\omega)$, where $\omega_0$ is the
energy of the unperturbed $0^-$ state, and $\omega$ is the energy of the TDA
phonon.  The result for $V^{PNC(2)}_{Dwy}$ consists of one isoscalar
contribution, given in Eq.~(4.9) of Ref.~\cite{john1}, and two isovector
contributions, given in Eqs.~(4.10) and (4.11) of that paper. The
renormalization amounts to a suppression of the isovector $0^-$
part $V^{PNC(2)}_{Dwy}$ by about a factor of about 3 with little change in
the isoscalar $0^-$ part (which was found to be much smaller for other
reasons).

Desplanques has extended the study by considering the $0^-$ resonances in the
random-phase approximation (RPA)~\cite{desplanques2} (see also
Ref.~\cite{flam1}).  This result shows that the multi-particle multi-hole
correlations are very important.  To get a nonvanishing contribution from the
collective excitations in RPA, it is necessary to include nonlocal terms in
the residual strong interaction.  The combined effect of these two
considerations is to undo most of the suppression arising from the
one-particle one-hole TDA correlations found in Ref.~\cite{john1}.
Desplanques found, essentially, that the doorway result changes by the
following substitutions,
\begin{eqnarray}
\label{renorm}
\omega_0/\omega \rightarrow \frac{1}{1+\lambda_p}, ~~~~~isoscalar \nonumber
\\ \omega_0/\omega \rightarrow \frac{1}{1+\lambda_p'}, ~~~~~isosvector ,
\end{eqnarray}
where $\lambda_p$ and $\lambda_p'$ are defined in terms of the Fermi-liquid
parameters as
\begin{eqnarray}
\label{momdep}
\lambda_p = {1\over 3}G_1 -{10\over 3} H_0+{4\over 3}H_1-{2\over 15}H_2
\nonumber
\\ \lambda_p' = {1\over 3}G'_1 -{10\over 3} H'_0+{4\over 
3}H'_1-{2\over 15}H'_2.
\end{eqnarray}
and where the effective mass of the nucleon $M^*$ is set to its free-space
value (the dominant momenta in the phonon are momenta above the Fermi
momentum $p_F$).  Using the Fermi-liquid parameters of Ref.~\cite{pqw}, we
calculate that $\lambda_p=-0.21$ and $\lambda_p'=0.27$.

We use here the RPA treatment~\cite{desplanques2} of the $O^-$ resonances
rather than the TDA treatment.  This is implemented by using
Eq.~(\ref{medium}) (see Appendix) with the replacements in Eq.~(\ref{renorm}).
The isovector interaction, which is the only significant piece to this doorway
result, continues to be suppressed in RPA, but the renormalization is about a
factor of two smaller than it was in Ref.~\cite{john1}.

\section{CALCULATIONS}

In this section we describe our calculation of $M$ and compare the
results to the values of $M$ deduced from scattering longitudinally
polarized neutrons from the target nuclei $^{238}$U and $^{104,105,106,108}$Pd.
These nuclei are good first cases to consider, because our calculation
is relatively straightforward for proton and neutron mid-shell occupancy, and
because the empirical values of the weak spreading width $\Gamma_w$ are rather
well determined~\cite{triple} for these cases.  The relationship between
$\Gamma_w$ and $M$ is most precisely stated if we affix a subscript $J$ on
$M$ to specify the total angular momentum of the CN states over which the
PNC interaction is averaged and introduce $D_J$, the theoretical level
spacing for s-wave levels of total angular momentum $J$ in the region of the
spectrum where the measurement has been performed.  Then,
\begin{equation}
\label{mm}
\Gamma_w= 2\pi|M_J|^2/D_J,
\end{equation}
where the label $J$ has been omitted from $\Gamma_w$ because of the expectation
that this quantity is insensitive to the details of the distribution of CN
levels.  The resulting values for $M_{Exp}$ are given in Table~III.  Also shown
in the table is the spacing $D_0$ of all s-wave levels, which is the same as
$D_J$ only for $J=0$ targets.  In the case of the daughter $^{106}$Pd there are
two relevant intermediate states, of $J=2$ and $3$, and in this case
the quantity $D_0$ is the combined level density for the two states.  To
determine value of $M_{Exp}$ in this case required making an assumption
about the relationship between $D_0$ and the two values of $D_J$.

We report the results of our calculation of $M$ based on the theory developed
in earlier sections of this paper in Table~IV. The contributions to $M$ arise
from two sources, the standard two-body parity violating interaction
$V^{PNC(2)}_{Std}$ (Sect.~III.A.2) and the doorway piece $V^{PNC(2)}_{Dwy}$
(Sect.~III.A.3).  To evaluate the matrix element $M$ corresponding
to $V^{PNC(2)}_{Dwy}$, we replace $\omega_0/\omega_{iv} \rightarrow 0.78$ in
Table~VII in accord with Eq.~(\ref{renorm}).  Details of our calculation 
of the matrix elements of $V^{PNC(2)}_{Dwy}$ and $V^{PNC(2)}_{Std}$ are given 
in the appendix. The values of $M$ corresponding to these two pieces of the 
interaction separately are given in the second and third columns of Table~IV, 
respectively.  In the case of the daughter $^{106}$Pd, there are two different 
sets of s-wave resonances
corresponding to $J=2$ and $J=3$ as discussed earlier; the individual values
of $M_J$ differ by less than 10\%, so we have given only the average value
$M$ in this case in Table~IV.  The small value of $M$ for $^{106}$Pd reflects 
smaller level spacing seen in Table~III and the near constancy of $\Gamma_{m}$ 
in the Pd isotopes.

The fourth column of Table~IV contains the value of $M$ corresponding to the
sum of {\it Dwy} and {\it Std}.  These results must be added in 
quadrature to obtain
the combined value of $M$.  Note that in doing this one must allow
for the possibility that the two pieces can interfere.  We
can easily verify from the values given in Table~VII that in practice there
is only a small interference between the  {\it Std} and {\it Dwy}
contributions.  We find
the interference is constructive for $A\sim 230$ but destructive for
$A\sim 100$.

In the last column of Table~IV we show the empirical
value of $M$ taken from Table~III.  We see that for $A\sim 100$ the 
experimental
values of $M$ fluctuate around the theoretical values.  
As the weak spreading widths are expected to be much more
stable than $M$ (the theoretical spreading widths are stable to within about
$\pm 10$ \% ), we compare in Table~V the theoretical and experimental
spreading widths in the two mass regions considered.  Given that the spreading
widths are
quadratic in $M$, we see from Table~VI that the $M$ in the mass $A\sim 230$
region is about a factor of 3 smaller than the empirical value and
in the mass $A\sim 100$ region about a factor of 1.7 smaller.  The agreement
between theory and experiment is remarkably close and encourages further
applications with the goal of refining the empirical determination of the
weak coupling parameters.

Earlier calculations~\cite{john3,flam,john1}  in the mass region $A\sim 230$
share some features in common with the present approach. 
Reference~\cite{john3}
described the nucleus as a chaotic Fermi gas, i.e. the basis states of the
nucleons were taken to be plane waves.  The framework of statistical
spectroscopy was used to calculate $M$ with the strength function
of the PNC interaction normalized to the strength function calculated in a
finite nucleus~\cite{tomsovic}.  When plane waves are used, the selection
rules are much weaker than with shell-model states, and for this reason both
$V^{PNC(2)}_{Std}$ and $V^{PNC(1)}_{Std}$ contributed {\it directly} to $M$.
In the present work $V^{PNC(1)}_{Std}$ contributes directly only if the
nuclear core is deformed (see Sect.~V.B).  Otherwise, $V^{PNC(1)}_{Std}$
contributes indirectly through spin-flip excitations.  As discussed in
Sect.~III.B.3, $V^{PNC(2)}_{Std}$
contributes directly because of the presence of opposite-parity intruder
states.  In Ref.~\cite{john3} $M$ was found to be several times larger than the
empirical value (see also Ref.~\cite{john1}), whereas in the present work it
appears to be several times smaller.  It would thus appear that the nuclear
matter treatment overestimates the result of statistical spectroscopy in
a finite nucleus for the contribution of our one-body PNC interaction for the
daughter $^{239}$U.

Reference~\cite{flam} employed a $0\hbar\omega$ model space like ours,
so $V^{PNC(1)}_{Std}$ contributed indirectly, through what can be thought of as
the perturbative piece of the doorway process.  The average over states was
defined by a Lorentzian function approximating the spreading width and
normalized to an empirical estimate of the number of ``principal" components
of the nuclear wave function.  Good agreement between measured and calculated
values of $M$ were reported.  In Ref.~\cite{flam} the indirect contribution
of $V^{PNC(1)}_{Std}$ dominated the contribution of $V^{PNC(2)}_{Std}$,
and the latter was neglected.  In our work, the matrix 
elements of $V^{PNC(2)}_{Std}$ are similar to those in Ref.~\cite{flam} 
(note that a different representation of the two-body PNC interaction was used 
in our work).  However, in contrast to Ref.~\cite{flam}, the matrix elements
of $V^{PNC(2)}_{Std}$ are comparable 
to the sizes of the matrix elements of our $V^{PNC(2)}_{Dwy}$, as can be 
inferred from Table~IV.  Although we have used different parametrizations of 
the effective strong interaction, we do not believe that this can fully 
account for the large differences we find.

It was suggested in Ref.~\cite{flam} that the contribution of 
$V^{PNC(1)}_{Std}$ to $M$, relative to that of $V^{PNC(2)}_{Std}$,
would follow an $A^{1/3}$ law.  We see from Table~IV
that the relative value of the doorway contribution does increase with $A$
as predicted in Ref.~\cite{flam}, but a more detailed estimate suggests
that it increases by more like a power of $A^{1/2}$.  The existence
of the different rate of increase of the standard and doorway
contributions shows that the A-dependence of the measurements may provide
an empirical means to determine the size of the doorway contribution.

\section{DISCUSSION}

In this section we will address the implementation of statistical spectroscopy
and various sources of uncertainty in the calculations of the previous
section.  Most of the uncertainties we discuss are amenable to further
analysis in large-basis shell-model studies, which would be valuable
to gain additional confidence in the theoretical results.

The theory of level densities and strength distributions within statistical
spectroscopy applies to individual systems and operators, and as such is not
inherently an ensemble theory.  The central limit theorem results and
correction expansions are all based on having many strongly interacting
particles dilutely spread among even more single-particle states.
The inputs required for all calculations are low-order operator moments.
If it were possible to calculate exact moments up to products of
approximately, say, six operators decomposed for the symmetries
involved, we would have an essentially "perfect" implementation of statistical
spectroscopy.  It would be extremely accurate and would work even far 
out in the
tails of the distributions.  Instead, we have made a number of practical
compromises in order to complete this first implementation for the
parity-violation problem.  Some of the compromises can and should be
eliminated in future, planned, improved calculations.

Most of the compromises we have chosen are motivated by the following
considerations:  {\it i}) the necessary angular momentum decomposed moments
cannot
be written down analytically; {\it ii}) even the non-decomposed moments
(scalar moments) are not worked out beyond a product of four operators--the
expressions for products of three or four operators, though they exist, are
almost prohibitively unwieldy; and, {\it iii}) the exact operators may not be
known in the model spaces--for example, the exact, residual, strong operator
in a model space appropriate for heavy nuclei.  We have addressed these points
by: {\it i}) treating the angular momentum as a statistical variable in the
spirit of Bethe's ansatz for the spin cutoff factor; {\it ii}) use partitioning
of the
moments to improve accuracy instead of incorporating polynomial corrections to
the lowest order central limit theorems--this alleviates much of the
need for higher moments; and, {\it iii}) supplement exact moments with ensemble
calculations and nuclear modeling--the idea being that in chaotically behaving
systems whole
classes of operators deviate very little from certain ensemble results
(subject to very few restrictions).  This also reduces the need for
higher-order moments.

To be more specific about ensemble results, we have in mind two-body random
operator ensembles which are a subclass of embedded Gaussian
ensembles~\cite{french2}.  The basic ensemble would have random k-body
matrix elements embedded in an m-particle space distributed over $N$
single-particle levels; two-body ($k=2$) is the most important one.  Some of
the moment simplifications that we are making in this paper, see the
comments after Eq.~(\ref{sdmoments}), are ensemble results.  Expectation
value relations such as
\begin{eqnarray}
\langle OH \rangle = \langle O \rangle \langle H \rangle \nonumber
\\ \langle O^2H^2 \rangle = \langle O^2 \rangle \langle H^2 \rangle ,
\label{cor}
\end{eqnarray}
which express a factoring approximation true in systems that behave
statistically, have been found to hold to a considerable accuracy (see
Refs.~\cite{french,tomsovic}).  Operators $O$ or their higher powers, which
effectively contain a piece proportional to $H$ (or a function of $H$), will
not satisfy relations of this kind.

In the following subsections, we address issues related to the nuclear
modeling we have done, specifically the size of the model space; deformation;
correlations between $\overline{O}$ and $\overline{H}$ including spurious
center of mass motion; and, the choice of weak coupling parameters.  We
critique each uncertainty below, in Sects.~V.A to V.D.  In Sect.~V.E we
discuss how the shell-model could be used to gain additional confidence in the
models.  All of the statistical spectroscopy compromises that were discussed
above could (and should) be incorporated into the shell model testing.

\subsection{Size of Model Space}

The physical motivation for choosing a $0\hbar\omega$ model space was
discussed in Sect.~III.  However, to establish that the optimal selection of
active orbits has been made, it must be verified that the amount of parity
mixing as determined by our statistical methods are stable under 
enlargements of
the model space.  For this purpose, we added a spherical $1f_{5/2}$ neutron
and $1g_{7/2}$ proton orbit to the model space for $^{239}$U.  Adding these
two orbitals, found that the calculation of $M$ decreased by about 10\% 
when the residual strong and weak interactions were
chosen as described above.  Note, according to the theory of Sect.~III, the
effective interactions should be changed as the model space is enlarged.
Although we have not made such compensating changes, we expect that the effect
is small for the small enlargements considered here.  Our results thus
indicate that the estimates based on the model space given in Table~I and
Table~II are sufficiently accurate for the purposes of comparing to
experiment, where the errors are greater than 10\%.

The reason for the stability is that the neutron and proton valence orbits
are not close to being either fully occupied or completely empty.  In the
opposite situation, choosing the optimal model
space becomes a more delicate issue.  There are a number of cases measured in
Ref.~\cite{triple} in which this is true for one of the shells (for example,
Nb and Cs), and in these cases the choice of the optimal model space requires
more effort.  The data show that the spreading width is anomalously small in
these cases, and it will be an interesting test of the theory to be able to
reproduce this trend.

\subsection{Deformation}

Deformation contributes in two possible ways:  if the nucleus has a
permanent deformation, the one-body PNC potential develops a nonspherical
component arising from the interaction of a valence nucleon with the deformed
core.  Otherwise, the two-body effective interaction may have a contribution
that arises as one of the nucleons interacts with a core nucleon, exciting
it and causing a core-deformation fluctuation.  A second nucleon may then
experience a long-range interaction with the first through the resulting core
polarization.  We estimate the effect of permanent deformation
in this section.

\subsubsection{Permanent Deformation}

The fact that the empirical and theoretical values of $M$ more nearly agree
for $A\sim 100$ than they do in the region of $A\sim 230$ could point to an
important contribution of permanent deformation, since $^{238}$U is a highly
deformed nucleus but Pd is not.  It is easy to show that there can be
non-vanishing
matrix elements of the deformed piece of $V^{PNC(1)}_{Std}$ for opposite parity
spherical orbits in which the $j$ differ by zero or {\it two} units (and for
opposite parity deformed orbits in which the z-projection quantum numbers
$\Omega$ are the same).  For the spherical orbits in Table~I there are
pairs of orbits for both protons and neutrons that satisfy this requirement and
have rather large matrix elements.  Thus, for our calculations in the
$A\sim 230$ region we will have direct contributions from $V^{PNC(1)}_{Std}$
even in the $0\hbar\omega$ model space, similar to the situation in
nuclear-matter~\cite{john3}.

We find an estimate for the contribution to $M$ arising from the deformation 
by using a one-body potential similar to that given in Eq.~(\ref{eq:16}) 
and using 
an Harmonic oscillator basis, replacing $\rho$ by a density having a 
deformed Woods-Saxon shape with radius $R$ and diffuseness $a$.
The contribution of the one-body PV potential arising from deformation is found
by expanding the density to lowest order in the deformation parameter 
$\beta_2$, leading to the following shape,
\begin{equation}
\label{dens}
\delta\rho(r)/\rho_0=-\frac{R\beta_2}{a}g(r)Y_{20}(\hat{r})
\end{equation}
where
\begin{equation}
\label{f}
g(r)=\frac{\rho (r)}{\rho_0} \frac{e^{(r-R)/a}}{1+e^{(r-R)/a}} .
\end{equation}
Taking matrix elements between the single-particle orbitals, we obtain
\begin{eqnarray}
\label{onebody}
<n_f \ell_f j_f m_i &|&\delta V^{PNC(1)} |n_i \ell_i j_i m_i >
 =  \frac{-im\omega_0 R\beta_2}{a} (C_0+C_1,C_0-C_1) \nonumber \\
&   & <n_f \ell_f j_f m_i
|{1\over 2}(g(r) \sigma \cdot p +\sigma \cdot p g(r) ) Y_{20}(r) 
|n_i \ell_i j_i m_i > ,
\end{eqnarray}
where at $\rho = \rho_0$, $C_0+C_1=3.53\cdot 10^{-8}$ and
$C_0-C_1=-3.15\cdot 10^{-9}$ for (proton,neutron) orbits, and where the
radial quantum number $n$ is the number nodes in the radial wave function,
$\ell$ is the orbital angular momentum, $j$ is the total angular momentum,
$m$ is the projection of the total angular momentum along the z-axis,
and
\begin{eqnarray}
\label{ob}
<n_f \ell_f j^{\pm}_f m_i |{1\over 2}(g(r) \sigma \cdot p & + &\sigma \cdot p g(r)) 
Y_{20}(r) |n_i \ell_i j_i
m_i > =  (-1)^{m_i-\frac{3}{2}} \sqrt{(2j_f+1)(2j_i+1)}\nonumber\\
\left\{
\begin{array}{ccc} j_i & 2 & j_f \\ \ell_f \pm 1 & \frac{1}{2} & \ell_i
\end{array}
\right\}
& &\sqrt{\frac{5(2\ell_i+1)}{4\pi}}
\left(
\begin{array}{ccc} j_i & 2 & j_f \\ m_i & 0 & -m_i
\end{array}
\right) 
\sqrt{2(\ell_f\pm 1)+1} \left(
\begin{array}{ccc} \ell_i & \ell_f \pm 1 & 2 \\ 0 & 0 & 0
\end{array}
\right) \nonumber\\
<n_f\ell_f &|&{1\over 2}(g(r) \sigma \cdot p +\sigma \cdot p g(r) ) 
|n_i\ell_i\pm 1> ,
\end{eqnarray}
with $j_f^{\pm}\equiv \ell_f \pm \frac{1}{2}$.

Matrix elements of Eq.~(\ref{onebody}) do not vanish in the model space
of Table~I.  There are two non-vanishing matrix elements, one for the pair
proton orbits ($h_{9/2},i_{13/2}$) and one for the pair of neutron
orbits ($j_{15/2}$,$i_{11/2}$).  The matrix element of $\delta V^{PNC(1)}$
of Eq.~(\ref{onebody}) is about 2.4 times smaller for neutrons
than for protons.   Averaging the square of Eq.~(\ref{onebody}) over 
the quantum
number $m_i$ (noting that there are $2j_<+1$ nonvanishing matrix elements),
we find that the magnitude of the average proton matrix element is 0.011 eV.

We have estimated $M$ corresponding to these values by normalizing to the
calculation in Ref.~\cite{tomsovic} along the lines explained in
Ref.~\cite{john3}, which gives
\begin{equation}
\label{msq}
M^2_{Def}(keV^2)=2.6 \alpha_p^2 ,
\end{equation}
where
\begin{equation}
\label{alpha}
\alpha_p^2=\frac{(1.2)^2}{A_v}\frac{Tr[(\delta V^{PNC(1)})^2]}{Tr[(U_2)^2]} .
\end{equation}
The factor of $1/A_v$ arises because the trace is taken over the
squares of the one- and two-body matrix elements within the model space;
the square of the matrix elements is weighted by the relative
number of valence nucleon pairs, and the factor of $1.2^2$ represents
the relative normalization of the strength function for a one-and two-body
operator.  The two-body interaction $U_2$ has
no diagonal matrix elements but is otherwise modeled after the surface
delta-function interaction~\cite{gb}.  We estimated in this fashion that
in $^{238}$U, $M_{Def}=0.004$ meV.  

Compared to the values of $M$ arising from
other sources in Table~IV, we see that the contribution of deformation to
$M$ is completely negligible.
A more quantitative calculation would entail a calculation using a deformed
model space of Nilsson orbits, which would require no new theoretical
development because statistical spectroscopy as formulated in Sect.~III.B
applies in either a spherical or deformed basis.

\subsubsection{Core Polarization}

Core polarization as described in the introduction to this section has been
studied phenomenologically, and its contribution to the effective
nucleon-nucleon interaction is known to act dominantly in $T=1$ and be largely
independent of spin~\cite{molinari}.  Just as in the case of the doorway
model~\cite{john1} the effective parity-violating interaction results from
the commutator with the residual strong interaction and ${\bf \sigma \cdot r}$.
This commutator does not vanish for the core-polarization piece of the
effective nucleon-nucleon interaction, and the resulting contribution to
the effective PNC interaction deserves quantitative study in the future.

\subsection{Correlations between $O$ and $H$}

There are several classes of correlations that we wish to mention.  One
arises when $O$ couples strongly to a mode of motion generated by $H$
lying predominantly outside the model space.  In such a case, one would
miss important physics by averaging $O$ over only those states within the model
space.  To deal with this situation properly, one must use effective
interaction theory and consider instead of $O$ its effective operator
counterpart $\overline{O}$.  The book keeping rules of the theory ensures
that corrections necessary to represent the effect of this correlation are
included in $\overline{O}$.

Such a correlation is quite important for parity violation,
where the isoscalar and isovector $0^-$ spin-flip resonances couple strongly
to the one-body piece of the parity violating interaction, so in this case
$O$ is identified with $V^{PNC(1)}_{Std}$.  In order to deal
with this situation in the context of parity violation, the doorway model was
introduced, as discussed in Sect.~III.A.3.  We showed that the effect of
the $0^-$ spin-flip excitations is quite important.

Yet another type of significant correlation occurs when
the operator $O$ and Hamiltonian $H$ commute with each other, or where both
commute with a third operator $R$.  Let us consider the latter case, so
that $H$ and $O$ may be characterized by the same quantum numbers $\Gamma$ that
are the good quantum numbers of $R$.  Values of the mean-square average of $O$
over distinct ensembles of states, each characterized by its own value of
$\Gamma$, would in general be different.  Thus, it would be misguided
to average $O$ over an ensemble of states with mixed $\Gamma$, since $H$ does
not produce such mixtures.  If a statistical ensemble of mixed $\Gamma$ were
formed, either because one was not aware of the symmetry $R$, or because
of some unfortunate approximation, this ensemble would fail to properly
describe the system.
 
One place where such a consideration is relevant is in the description of
spurious center-of-mass motion.  Generally, the eigenstates of $H$ may be
grouped into sets distinguished by quantum numbers specifying the
center-of-mass motion.  Suppose that each such set is spanned by distinct
individual-particle basis states.  Then, any mean-square average of $O$
corresponding to one set would differ from that of any other, because the
corresponding statistical ensembles are different.  For example, if one
compares two ensemble averages of $O$, one in which the center-of-mass motion
is pure, and one in which it is a mixture of several modes, one may find
different results.  Fortunately, this problem turns out not to be a serious
one for us.  As can be seen from Tables~I and II, our $0\hbar\omega$ basis
does not permit any spurious center-of-mass excitations.  Thus, all matrix
elements of $V^{PNC(2)}$ are completely nonspurious.  As is generally the case
in applying perturbation theory to include excitations outside the model
space, some level of spuriosity is unavoidably introduced in calculating
$V^{PNC(1)}$.  However, the nature of statistical calculations, where the
observable is a mean-squared matrix element insensitive to relative phases,
considerably reduces the magnitude of the problem compared to the standard
shell-model case.

\subsection{Choice of Weak Coupling Parameters}

It is important to bear in mind that  the relative sizes of $M$ corresponding
to $V^{PNC(2)}_{Dwy}$ and $V^{PNC(2)}_{Std}$ depend upon the choice of weak
coupling constants.  The values of these coupling constants are not certain,
so the relative sizes of $M$ for the two contributions may actually be
different (perhaps substantially so) than the values reported in Tables~III-V
when more refined choices of the coupling constants are made.
As we have indicated, we
use the DDH parameter set for our calculations given in these tables.  A more
detailed breakdown of the operators
involved as well as a careful statistical study of the additional empirical
information available from TRIPLE and from other sources should allow
extraction of a consistent set of values for $F_\pi$ and $F_0$ from
experiment.  The application of our statistical theory to the remaining nuclei
of the TRIPLE data set requires a more elaborate study because, as we have
remarked, in cases not examined here the valence shells contain fewer nucleons
and are therefore more sensitive to details requiring further refinement,
such as the size of the model space and possibly the single particle energies.

Note that the size of the {\it Dwy} contribution in our calculations is
determined predominantly by our choice of the weak pion-nucleon coupling
constant.  If, instead of the DDH value for this coupling parameter, we would
rely instead on empirical values inferred from $^{18}$F~\cite{adelberger},
{\it Dwy} would be considerably smaller.  Likewise, the value we found for the
{\it Std} component of the force may change substantially.  For example, if
the measured anapole moment of $^{133}$Cs is taken to determine the weak $\pi$
and $\rho$ coupling constants, with $F_\pi$ constrained by the measurement in
$^{18}$F, the value of the weak $\rho$ coupling constant would be
substantially larger~\cite{wb}, as would our value of the {\it Std}
contribution in Table~IV.  The interplay between the role of the  {\it Std}
and {\it Dwy} contribution clearly depends on the region in $F_\pi$ - $F_0$
space of interest, the details of which will be reported elsewhere.

\subsection{Shell Model Tests}

There are a number of useful tests of the validity of the statistical model for
parity violation that could be carried out within large-basis standard shell
model
calculations.  We give one example:  how to model parity mixing, as manifested
in $M$, as it develops between states within the $0\hbar\omega$ model space
and those lying outside it.  We have assumed in this paper that the mixing
occurs through
the doorway mechanism as modified by Desplanques~\cite{desplanques2}.
The assumption can be evaluated in a toy model (based on large but finite
spaces and schematic interactions) as follows:  On the one hand, the chosen
interaction
may be diagonalized numerically (thus obtaining a model-exact result) using a
large-basis shell model code.  This same problem may then be examined 
in a model
space applying a version of effective interaction theory.  Various questions
may be answered by comparing results obtained from the exact diagonalization
and from effective interaction theory.  Limiting cases would be instructive to
consider.

For example, if excited states are initially well separated from the
model space by a large gap, the leading terms in the perturbative expansions
discussed in Sect.~III will dominate.  In the opposite limit of a very small
gap, which corresponds to the well defined case of statistical mixing in the
full space, the calculation again becomes simple.  The more difficult question
of theoretical interest is how to handle the intermediate cases, where the gap
is neither small nor large.  It is in this regime where the doorway model has
been conjectured to be applicable.  It would be interesting to: (1) confirm the
doorway model in specific numerical cases.  It may even be possible to motivate
extensions of it valid over larger gap sizes by making selective summations
over specific classes of diagrams to account for multiple p-h excitations;
(2) investigate the relative importance of effective three-body PNC
interactions, which have so far been neglected in all theoretical studies.
In the most favorable outcome, the doorway result we have already evaluated
describes this mixing over gaps comparable to shell spacings.
Well-posed questions in such a model could lead to a better understanding of
the theoretical uncertainties in applying statistical spectroscopy to PNC.

\section{SUMMARY AND CONCLUSION}

In this paper, we have made the first extension of statistical spectroscopy to
the case of parity violation and the weak spreading width in nuclei as
measured by the TRIPLE collaboration~\cite{triple}.  We have emphasized that
statistical strength function methods are advantageous in this case because
the observables are expressed in terms of matrix elements of the effective PNC
interaction averaged over {\it squares} of wave function components in a basis
of independent-particle model states.  We have also stressed that the
corresponding theoretical results are less sensitive to theoretical
uncertainties in familiar shell-model calculations even
though the nuclear wave functions are very complicated.

In extending statistical spectroscopy to parity violation, we find that it is
essential to include corrections to the underlying parity violating
interaction to account for nonstatistical correlations.  Of particular
importance are the spin-flip correlations between states separated by
$n\hbar\omega$, with $n\ge 1$.  In this regard, effective
interaction theory is particularly useful, and the $n \hbar\omega$ corrections
have been incorporated through an effective operator.  In calculating the
effective operator of the PNC interaction we have implemented in the RPA
approximation suggested by Desplanques~\cite{desplanques2}.

We have evaluated the weak spreading width for $A\sim 230$ in the $^{238}$U
target and for $A\sim 100$ in the $^{104,105,106,108}$Pd targets
using the standard estimates for the weak coupling parameters of Desplanques,
Donoghue, and Holstein~\cite{ddh}.  The theoretical results are in
qualitative agreement with the experimental results:  the measured $M$ in the
Pd isotopes differs by about a factor of 1.7 and in U by a factor of 3.  This
observation supports the hope that ambiguities present in the values of the
underlying weak meson-nucleon coupling parameters may be settled by using the
weak spreading width measured
via neutron scattering from compound nuclei~\cite{triple}.

We have investigated the sensitivity or our results to various improvements
in the theory, such as the sensitivity to enlargements of the model space
and to permanent deformation.  Permanent deformation seems to be a small
correction, and our estimates indicate that results for M are stable to 
10\% accuracy under enlargements of the space.  
The reason for the stability is that these nuclei
are in regions of the periodic table where large numbers of neutrons and
protons occupy the shells.  When, on the other hand, shells are nearly empty
(or nearly full), choosing the optimal model space becomes a more delicate
issue.  We have indicated that the explanation of the anomalously small
spreading width in the Nb and Cs cases~\cite{triple} will constitute an
interesting test of the theory.  Because of the extended space required,
these calculations require considerably more effort and have therefore
not been considered in this paper.  We have suggested additional toy-model
calculations that might provide insight and improve confidence in the
theoretical results.

\acknowledgments The work of MBJ and JDB was supported in part by the U.S.
Department of Energy under contract W-7405-ENG-36 and that of ST by
the National Science Foundation grant PHY-9800106.  ST expresses his gratitude
for the
hospitality extended by the Physics Division of Los Alamos National Laboratory.

\newpage

\appendix {{\bf Appendix:  Matrix Elements of Effective Parity-Violating
Interaction}}

In this Appendix we give explicit expressions for the effective two-body
PNC interaction and their matrix elements in an harmonic oscillator basis
that have been used for our numerical calculations.

We may represent $V^{PNC}$ in general as
\begin{eqnarray}
\label{freespace}
V^{PNC}=\sum_\alpha V_{\alpha}^{PNC} ,
\end{eqnarray}
where $V_{\alpha}^{PNC}$ are terms depending on spin operators
${\bf \theta}^{S(\alpha )}$ and isospin operators $\theta^{T(\alpha )}$.
The free-space PNC interaction in the meson-exchange model is given in
Table~VI.  Although there have been numerous estimates of
the strengths of the
weak meson-nucleon coupling constants in the literature, for our numerical
work we use the DDH "best" values of $F_{\pi}$ and $F_0$ (corresponding
to the values $f_\pi=0.454~10^{-6}$ and $h^0_\rho = -1.14~10^{-6}$).  As we
stated earlier, we omit the other coupling constants, since they are quite
small in the DDH analysis.

The interplay between the two-body strong and weak interaction is
approximately described~\cite{adelberger} by the two-body correlation function
$f(r)$ given
by Ref.~\cite{miller}.  The modification in $V^{PNC(2)}_{Std}$ arising from
the short-range correlation function occurs by multiplying the initial
and final nuclear wave function by $1+f(r)$, where $f(r)$ describes the
suppression of the relative wave function at short distances due to the action
of the nucleon-nucleon interaction.  Otherwise, the representation of
$V^{PNC(2)}_{Std}$ is the same as that for the free-space interaction as given
in Eq.~(\ref{freespace}) and Table~VI.

Our representation of $V^{PNC(2)}_{Dwy}$ is
\begin{eqnarray}
\label{medium}
V^{PNC(2)}_{Dwy}=\sum_{\alpha=7,9,11,13} {\bf \theta}^{S(\alpha )} \cdot
{\bf v}^{\alpha}_1(r)V^{\alpha}_{0}(R)\theta^{T(\alpha )}
+\sum_{\alpha=15,17} {\bf \theta}^{S(\alpha )} \cdot
{\bf V}^{\alpha}_1(R) v^{\alpha}_{0}(r)\theta^{T(\alpha )} ,
\end{eqnarray}
where a dependence on $R$, the center-of-mass position of the two
nucleons, is now possible because of the presence of a third body 
(the nucleus).  The definition of the various quantities is given in Table~VII.
The pieces associated with the isovector $0^-$ resonance are proportional to
$C_1$, which is dominated by
$F_\pi$.  These, in particular Eq.~(4.10) of Ref.~\cite{john1}, are much more
important than the
isoscalar piece associated with the isoscalar $0^-$ resonance, which is
proportional to $C_0$ and dominated by $F_0$ (the numerical values of 
$C_\pi$ and $C_{\rho}$ in Eq.~(4.10) of Ref.~\cite{john1} are misquoted 
there and should rather be identified with the values of $c^s_\rho$ and 
$c^v_\pi$ given in Eq.~(\ref{eq:17}), given next).  These constants are given 
in the local density approximation by
\begin{eqnarray}
C_0=c^s_\rho F_0 + c^s_\pi F_\pi {N_c-Z_c\over A_c} \nonumber
\\C_1=c^v_\pi  F_\pi+c^v_\rho F_0 {N_c-Z_c\over A_c}  ,
\label{eq:17}
\end{eqnarray}
where $c^s_\rho=7.61~10^{-3}$, $c^s_\pi=1.74~10^{-2}$, $c^v_\pi=1.74~10^{-2}$, 
and $c^v_\rho=1.3~10^{-3}$,
and where $N_c$ and $Z_c$ are the neutron and proton numbers of the core
($A_c=N_c+Z_c$).  In our evaluation of $V^{PNC(2)}_{Dwy}$ in this paper,
we have taken $\rho = \rho_0$.

Next, we give an explicit expression for the matrix elements
of $V^{PNC(2)}_{Std}$ and $V^{PNC(2)}_{Dwy}$.  We express the matrix element
of the effective interaction $\overline{V}^{PNC(2)}$
in second-quantized notation as
\begin{eqnarray}
\label{twobody}
\overline{V}^{PNC(2)}=\frac{1}{4} \sum_{\beta j \beta^{\prime} j^{\prime} J}
(2J+1)^{\frac{1}{2}}
<\beta_1 j_1; \beta_2 j_2 ||\tilde{V}^{PNC(2)}_J||
\beta_1^{\prime} j_1^{\prime}; \beta_2^{\prime} j_2^{\prime} >
\left\{ (a^{\dagger}_{\beta_1 j_1} a^{\dagger}_{\beta_2 j_2})^J
(\tilde{a}_{\beta_2^{\prime}  j_2^{\prime} } \tilde{a}_{\beta_1^{\prime}
  j_1^{\prime} })^J  \right\}^0_0 ,
\end{eqnarray}
where the sum runs over all quantum numbers $(\beta,j)$, where $\beta$
corresponds to the radial quantum number $n$, the orbital angular momentum,
the spin, and the isospin of the nucleon.  Here $a^{\dagger}_{\beta_1 j_1}$
is a creation operator, and $\tilde{a}_{\beta_1  j_1}$ is related to the
corresponding annihilation operator by $\tilde{a}_{jm}=(-1)^{j+m}a_{j,m}$.
The density matrix corresponds to initial and final pairs of particles
(of total angular momentum $J$) coupled to $\Delta J=0$.  The matrix element
is antisymmetrized
\begin{eqnarray}
<\beta_1 j_1; \beta_2 j_2 ||\tilde{V}^{PNC(2)}_J||
\beta_1^{\prime} j_1^{\prime}; \beta_2^{\prime} j_2^{\prime} >
&=&<\beta_1 j_1; \beta_2 j_2 ||\overline{V}^{PNC(2)}_J||
\beta_1^{\prime} j_1^{\prime}; \beta_2^{\prime} j_2^{\prime} > \nonumber\\
& & -(-1)^{j_1+j_2-J}<\beta_2 j_2; \beta_1 j_1 ||\overline{V}^{PNC(2)}_J||
\beta_1^{\prime} j_1^{\prime}; \beta_2^{\prime} j_2^{\prime} > ,
\label{anti}
\end{eqnarray}
and reduced in angular momentum only (we calculate
separately matrix elements for neutron pairs, proton pairs, and neutron-proton
pairs).

A general expression for $\overline{V}^{PNC(2)}_J$ is
\begin{eqnarray}
& & \Big\langle
\beta_1 j_1; \beta_2 j_2 ||\overline{V}^{PNC(2)}_J||
\beta_1^{\prime} j_1^{\prime}; \beta_2^{\prime} j_2^{\prime}
\Big\rangle = \sum_{\alpha}\left\langle N_{\beta_1} N_{\beta_2} |
T^{(\alpha_T)} |N_{\beta_{1^{\prime}}}N_{\beta_{2^{\prime}}}\right\rangle
\sum_{n\ell\atop NL}\sum_{n^{\prime}\ell^{\prime}\atop N^{\prime}L^{\prime}}
\sum_{S{\cal L}}\sum_{S^{\prime}{\cal L}^{\prime}} \nonumber\\
& & \qquad *
\left[\left(2j_{1}+1\right)\left(2j_{2}+1\right)\left(2j_{1}^{\prime}+1\right)
\left(2j_{2}^{\prime}+1\right)\left(2{\cal L}+1\right)^{2}\left(2S+1\right)^{2}
\left(2{\cal 
L}^{\prime}+1\right)\left(2S^{\prime}+1\right)\right]^{1/2}\nonumber\\
& & \qquad * \left\{ \begin{array}{ccc} \ell_{1} & \ell_{2} & {\cal 
L}\\ {1\over 2} &
{1\over 2} & S\\
j_{1} & j_{2} & J \end{array}\right\}
\left\{ \begin{array}{ccc} \ell_{1}^{\prime} & \ell_{2}^{\prime} & 
{\cal L}^{\prime}\\
{1\over 2} & {1\over 2} & S^{\prime}\\ j_{1}^{\prime} & j_{2}^{\prime} & J
\end{array}\right\}
\left\langle n^{\prime}\ell^{\prime},N^{\prime}L^{\prime},{\cal
L}^{\prime}\big\vert
n_{1}^{\prime}\ell_{1}^{\prime},n_{2}^{\prime}\ell_{2}^{\prime},{\cal
L}^{\prime}
\right\rangle\nonumber\\
& & \qquad * \left\langle n\ell,NL,{\cal L}\big\vert n_{1}\ell_{1},
n_{2}\ell_{2},{\cal L}\right\rangle
(-)^{J+{\cal L}^{\prime}+S}\left\{\begin{array}{ccc} {\cal L} & {\cal 
L}^{\prime}
& 1\\ S^{\prime} & S & J \end{array}\right\} \left[ 3(2\ell 
+1)(2L+1)\right]^{1/2}
\nonumber\\
& & \qquad* \left\{\begin{array}{ccc} {\cal L} & {\cal L}^{\prime} & 1\\ \ell &
\ell^{\prime} &
k_{1}\\ L & L^{\prime} & k_{2} \end{array}\right\}\left\langle S\left\|
S_{\lambda}^{S(\alpha_S)}\right\| S^{\prime}\right\rangle
\left\langle n\ell\left\| v_{k_{1}}^{\alpha}\right\|n^{\prime}
  \ell^{\prime}\right\rangle
\left\langle N L\left\| V_{k_{2}}^{\alpha}\right\| N^{\prime}
L^{\prime}\right\rangle\,\, ,
\end{eqnarray}
where the functional dependence of $\alpha_S$, $\alpha_T$, $k_1$, and
$k_2$ on $\alpha$ as well as the values of $v^{\alpha}_{k1}$,
$V^{\alpha}_{k2}$, $S^{(\alpha_S)}$, and $T^{(\alpha_T)}$
are given in Tables~VI and VII for the standard and doorway pieces of the
effective interaction, and we have made a Moshinsky transformation
to the relative and
center-of-mass coordinates ($\left\langle n\ell,NL,{\cal L}\big\vert n_{1}
\ell_{1},n_{2}\ell_{2},{\cal L}\right\rangle$ is a Moshinsky
bracket~\cite{brm}).  This is a generalization of the results given in
Ref.~\cite{castel}.  Our angular momentum conventions follow Brink and
Satchler~\cite{bs}.  We have checked our results against various special
cases, and two codes have been written independently to check the calculation
of $V^{PNC(2)}_{Std}$.  The $V^{\Gamma}_{rstu}$ of Eq.~(\ref{moments}) is equal
to the $\tilde{V}^{PNC(2)}_J$ in Eq.~(\ref{twobody}) multiplied by
$\sqrt{2}$ when the quantum numbers of particles 1 or 2 are equal in the
initial or final states.

\begin{table}
\caption {0$\hbar\omega$ model space for mass region $A\sim 230$.  Note that
we assume neutron (N) closure at $N=126$ and proton (P) closure at $Z=82$.  The
single-particle energies are given for the daughter $^{239}$U.}

\begin{tabular}{cccc}
Particle & Orbit  & (n,$\ell$) & Parity \\
\hline
P & h$_{9/2}$& (0,5) & -  \\
P & i$_{13/2}$& (0,6) & +  \\
P & f$_{7/2}$& (1,3) & -  \\
N & i$_{11/2}$& (0,6)& +  \\
N & j$_{15/2}$& (0,7)& -  \\
N & g$_{9/2}$& (1,4)& +  \\
N & d$_{5/2}$& (2,2)& +  \\
\end{tabular}
\end{table}

\begin{table}
\caption{0$\hbar\omega$ model space for mass region $A\sim 100$.  Note that
we assume neutron (N) closure at $N=50$ and proton (P) closure at $Z=28$.  The 
single-particle energies are given for the daughter $^{107}$Pd.}

\begin{tabular}{c c c c c}
Particle & Orbit  & (n,$\ell$) & Parity & Orbital Energy (MeV) \\
\hline
P & g$_{9/2}$& (0,4) & + & -7.94 \\
P & p$_{1/2}$& (1,1) & - & -9.04 \\
P & f$_{5/2}$& (0,3) & - & -11.2 \\
P & p$_{3/2}$& (1,1) & - & -10.7 \\
N & d$_{5/2}$& (1,2) & + & -8.44 \\
N & g$_{7/2}$& (0,4) & + & -7.34 \\
N & s$_{1/2}$& (2,0) & + & -6.63 \\
N & d$_{3/2}$& (1,2) & + & -5.96 \\
N & h$_{11/2}$& (0,5) & - & -4.85 \\
\end{tabular}
\end{table}

\begin{table}
\caption {The empirical weak spreading width $\Gamma_w^{Exp}$ [3] (preliminary)
for the daughters $^{239}$U and $^{105,106,107,109}$Pd.  Also shown are the
neutron number N, the proton number Z, the theoretical
level spacings $D$, and the value of $M$ calculated according to
Eq.~(\ref{mm}).}

\begin{tabular}{c c c c c}
Nucleus & (N,Z) & $\Gamma_w^{Exp}$ (10$^7$ eV) & D$_0$ (eV) & $M_{Exp}$
(meV) \\
\hline
$^{239}$U & (146,92) & 1.35$^{+0.97}_{-0.64} $ & 20.75 (20.9) &
$0.67^{+0.24}_{-0.16}$
 \\ $^{105}$Pd & (59,46)& 1.40$^{+6.91}_{-0.99} $ & 165 (165$\pm$ 61) &
$2.2^{+2.4}_{-0.9}$
 \\ $^{106}$Pd & (60,46)& 0.086$^{+0.098}_{-0.042} $ & 13.1 (13.3$\pm$ 0.7) &
$0.20^{+0.10}_{-0.07}$
 \\ $^{107}$Pd & (61,46)& 0.18$^{+0.91}_{-0.15} $ & 159 (159$\pm$ 24) &
$0.79^{+0.88}_{-0.36}$                                                
 \\ $^{109}$Pd & (63,46)& 0.81$^{+4.45}_{-0.62} $ & 159 (159$\pm$ 24) &
$1.6^{+2.0}_{-0.7}$                                                
 \\                       
\end{tabular}
\end{table}

\begin{table}
\caption {Theoretical values of $M$ for the effective parity-violating
interaction.  Contributions are shown separately
for the standard ( {\it Std}) and doorway ({\it Dwy}) pieces of the
two-body interaction.
A comparison of the experimental value of M given in Table~III is also shown.}
 
\begin{tabular}{c c c c c}
Nucleus & $M_{Std}$ (meV)  & $M_{Dwy}$ (meV)  & $M_{Std+Dwy}$
(meV) & $M_{Exp}$ (meV) \\
\hline
$^{239}$U & 0.116 & 0.177 & 0.218 & $0.67^{+0.24}_{-0.16}$
\\
$^{105}$Pd & 0.70  & 0.79  & 1.03   & $2.20^{+2.4}_{-0.9}$
\\
$^{106}$Pd & 0.304 & 0.357 & 0.44   & $0.20^{+0.10}_{-0.07}$
\\
$^{107}$Pd & 0.698 & 0.728 & 0.968 & $0.79^{+0.88}_{-0.36}$
\\
$^{109}$Pd & 0.73  & 0.72 & 0.97  & $1.6^{+2.0}_{-0.7}$
\\
\end{tabular}
\end{table}

\begin{table}
\caption {Comparison of theoretical weak spreading width to experiment.
The value of $\Gamma_w^{Exp}$ for Pd is obtained by statistically combining the
results for all
values given in Table~III; the theoretical value of $\Gamma_w^{The}$ is, to
within a few percent, the same for all Pd isotopes.}
 
\begin{tabular}{c c c }
Nucleus & $\Gamma_w^{The}$(10$^7$ eV)  & $\Gamma_w^{Exp}$(10$^7$ eV) \\
\hline
$^{239}$U & 0.143 &  1.35$^{+0.97}_{-0.64} $ \\
Pd & 0.40 & $1.73^{+1.65}_{-0.84} $ \\
\end{tabular}
\end{table}

\begin{table}

\caption {Free-space parity violating interaction of Desplanques, Donaghue,
and Holstein, Ref.~[10].}

\hbox to\hsize{\hrulefill}%
\vskip -18pt
\hbox to\hsize{\hrulefill}%
\noindent (a) Definition of $V_{\alpha}^{PNC}\equiv
\theta^{T(\alpha)}${\bfslant\char'022}$^{S(\alpha)}\cdot{\bf v}^{\alpha}$ for
the standard PNC interaction.
\vskip -3pt
\hbox to\hsize{\hrulefill}%
\def\onea#1#2#3#4#5#6{\hbox{
\hbox to 0.5truein{\hfill #1\hfill}%
\hbox to 0.5truein{\hfill #2\hfill}%
\hbox to 0.5truein{\hfill #3\hfill}%
\hbox to 2.0truein{\hfill #4\hfill}%
\hbox to 1.25truein{\hfill #5\hfill}%
\hbox to 1.75truein{#6\hfil}}%
}
\onea{$\alpha$}{$\alpha_{S}$}{$\alpha_{T}$}{$\theta^{T(\alpha)}=T^{(\alpha
_{T})}$}
{$\theta^{S(\alpha)}={\bf S}^{(\alpha_{S})}$}{\hfill $v^{\alpha}$\hfill}%
\vskip -3pt
\hbox to\hsize{\hrulefill}%
\onea {1}{2}{3}{$t_{10}^{(3)}$}{${\bf S}^{(2)}$}{${\sqrt{2}\over M}\left(
F_{\pi}{\bf u}_{\pi}^{(-)}+H_{1}{\bf u}_{\rho}^{(-)}\right)$}%
\medskip
\onea {2}{1}{4}
{$F_{0}t_{00}^{(4)}-{1\over 2\sqrt{3}}F_{1} t_{10}^{(2)}
-{1\over 2\sqrt{3}}F_{2}t_{20}^{(6)}$}
{${\bf S}^{(1)}$}
{$-{\sqrt{3}\over M}{\bf u}_{\rho}^{(+)}$}%
\medskip
\onea {3}{3}{4}{$F_{0}t_{00}^{(4)}-{1\over 2\sqrt{3}}F_{1}
t_{10}^{(2)}-{1\over 2\sqrt{3}}F_{2}t_{20}^{(6)}$}{${\bf S}^{(3)}$}
{$-{\sqrt{6}\over M}(1+\mu_{V}){\bf u}_{\rho}^{(-)}$}%
\medskip
\onea {4}{1}{2}{$G_{0}t_{00}^{(0)}+{1\over 2}G_{1}t_{10}^{(2)}$}{${\bf
S}^{(1)}$}
{${1\over M}{\bf u}_{\omega}^{(+)}$}%
\medskip
\onea {5}{3}{2}{$G_{0}t_{00}^{(0)}+{1\over 2}G_{1}t_{10}^{(2)}$}{${\bf
S}^{(3)}$}
{${\sqrt{2}\over M}(1+\mu_{S}){\bf u}_{\omega}^{(-)}$}%
\medskip
\onea {6}{2}{1}{$t_{10}^{(1)}$}{${\bf S}^{(2)}$}{${1\over 2M}
\left(G_{1}{\bf u}_{\omega}^{(+)}-F_{1}{\bf u}_{\rho}^{(+)}
\right)$}%
\hbox to\hsize{\hrulefill}%
\newpage
\hbox to\hsize{\hrulefill}%
\noindent (b) Definition of operators in Table VI(a).  Note that the momentum
operator ${\bf p}$ is defined as twice the relative momentum,
${\bf p}={\bf p}_1-{\bf p}_2$.
\vskip -3pt
\hbox to\hsize{\hrulefill}%
\def\oneb#1#2#3#4{\hbox{
\hbox to 0.9truein{#1\hfill}%
\hbox to 2.35truein{#2\hfill}%
\hbox to 0.9truein{#3\hfill}%
\hbox to 2.35truein{#4\hfill}}%
}
\oneb { }{$\qquad$Operator}{ }{$\qquad$Operator}%
\vskip -3pt
\hbox to\hsize{\hrulefill}%
\oneb {Spin:}{ }{Isospin:}{ }%
\oneb {$\quad${\bf S}$^{(1)}$}{{\bfslant\char'033}${(1)}-$
{\bfslant\char'033}${(2)}$}
{$\quad t_{00}^{(0)}$}{1}%
\oneb {$\quad${\bf S}$^{(2)}$}{{\bfslant\char'033}${(1)}+$
{\bfslant\char'033}${(2)}$}
{$\quad t_{10}^{(1)}$}{$\tau_{z}(1)-\tau_{z}(2)$}%
\oneb {$\quad${\bf S}$^{(3)}$}{$i$
{\bfslant\char'033}${(1)}
\times$
{\bfslant\char'033}${(2)}
/\sqrt{2}$}
{$\quad t_{10}^{(2)}$}{$\tau_{z}(1)+\tau_{z}(2)$}%
\oneb {Space:}{ }
{$\quad t_{10}^{(3)}$}{$i\big[${\bfslant\char'034}$(1)$ $\times$
{\bfslant\char'034}$(2)
\big]_{z}/\sqrt{2}$}%
\oneb {$\quad${\bf u}${(r)}^{(-)}$}{$\left[{\bf p},u_{0}\right]$}
{$\quad t_{00}^{(4)}$}{$-${\bfslant\char'034}$(1)$ $\cdot$ 
{\bfslant\char'034}$(2)
/\sqrt{3}$}%
\oneb {$\quad${\bf u}${(r)}^{(+)}$}{$\left\{{\bf p},u_{0}\right\}$}
{$\quad t_{20}^{(6)}$}{$\big[3\tau_{z}(1)\tau_{z}(2)-$ {\bfslant\char'034}(1)
$\cdot$ {\bfslant\char'034}$(2)\big]/\sqrt{6}$}%
\oneb {$\quad u_{0}(r)$}{$e^{-mr/4\pi r}$}{ }{ }%
\vskip -3pt
\hbox to\hsize{\hrulefill}%
\newpage
\hbox to\hsize{\hrulefill}%
\end{table}

\begin{table}
\caption {Effective interaction arising from the doorway
contribution as formulated in Ref.~[18].  Some of the notation is
defined in Table~VI.  Note that the doorway contribution is defined for
odd values of $\alpha$ only.}

\begin{center}
\begin{tabular}{ccccccccc}

$\alpha$ & $\alpha_{S}$ & $\alpha_{T}$ & $k_{1}$ & $k_{2}$ &
$\theta^{T(\alpha)}=T^{(\alpha
_{T})}$ & $\theta^{S(\alpha)}={\bf S}^{(\alpha_{S})}$
  & $v^{\alpha}_{k1}(r)$ (fm$^{-1}$) &
$V^{\alpha}_{k2}(R)$\\
\hline
$\phantom{0}$7 & 3 & 4 & 1 & 0 & ${-\sqrt{6}M\mu\omega_0
\over \omega_{is}\hbar c}\lambda_{11}
C_{0}t_{00}^{(4)}$ & $S^{(3)}$ & ${i\mu\over 4\pi}e^{-\mu r}\hat{r}$
& 1 \\
$\phantom{0}$9 & 3 & 2 & 1 & 0 & ${\sqrt{2}M\mu\omega_0\over \omega_{is}
\hbar c}\lambda_{10}
C_{0}t_{00}^{(0)}$ & $S^{(3)}$ & ${i\mu\over 4\pi}e^{-\mu r}\hat{r}$
& 1 \\
11 & 2 & 3 & 1 & 0 & ${\sqrt{2}M\mu\omega_0\over 2\omega_{iv}
\hbar c}(\lambda_{01}+\lambda_{11})
C_{1}t_{10}^{(3)}$ & $S^{(2)}$ & ${i\mu\over 4\pi}e^{-\mu r}\hat{r}$
& 1 \\
13 & 3 & 4 & 1 & 0 & ${\sqrt{2}M\mu\omega_0
\over 2\omega_{iv}\hbar c}(\lambda_{10}+\lambda_{11})
C_{1}t_{10}^{(2)}$ & $S^{(3)}$ & ${i\mu\over 4\pi}e^{-\mu r}\hat{r}$
& 1 \\
15 & 1 & 3 & 0 & 1 & ${\sqrt{2}M\mu\omega_0\over \omega_{iv}
\hbar c}(\lambda_{01}-\lambda_{11})
C_{1}t_{10}^{(3)}$ & $S^{(1)}$ & ${i\over 4\pi r}e^{-\mu r}$
& $\mu\vec{R}$ \\
17 & 3 & 1 & 0 & 1 & ${\sqrt{2}M\mu\omega_0\over \omega_{iv}
\hbar c}(\lambda_{10}-\lambda_{11})
C_{1}t_{10}^{(1)}$ & $S^{(3)}$ & ${i\over 4\pi r}e^{-\mu r}$
& $\mu\vec{R}$ \\
\end{tabular}
\begin{tabular}{lll}
$\lambda_{00}$ = --53.9 MeV-fm$^{3}$ & $\lambda_{11}$ = 239 MeV-fm$^{3}$& \\
$\lambda_{01}$ = 200 MeV-fm$^{3}$ &$\lambda_{10}$ = 59.8 MeV-fm$^{3}$ & \\
$M$ = 4.758 fm$^{-1}$ (nucleon mass)  & $\mu$ = 3.897 fm$^{-1}$ ($\rho$
meson mass) & \\
\end{tabular}

\end{center}
\end{table}

\end{document}